\documentclass[astrosymb,twocolumn]{aastex631}

\begin{document}


\newcommand{\figs}{Figs.}
\newcommand{\Fig}{Fig.}
\newcommand{\EFig}{Extended Data Fig.}
\newcommand{\Figs}{Figs.}
\newcommand{\sect}{Section}
\newcommand{\sects}{Sections}
\newcommand{\Sect}{Section}
\newcommand{\Sects}{Sections}
\newcommand{\tab}{Table}
\newcommand{\tabs}{Tables}
\newcommand{\Tab}{Table}
\newcommand{\Tabs}{Tables}
\newcommand{\eqn}{equation}
\newcommand{\eqns}{equations}
\newcommand{\Eqn}{Equation}
\newcommand{\Eqns}{Equations}
\newcommand{\etal}{et~al.}

\newcommand{\src}{GLEAM-X\,J\,0704\ensuremath{-}37}
\newcommand{\GLXlong}{GLEAM-X\,J\,162759.5\ensuremath{-}523504.3}
\newcommand{\GLX}{GLEAM-X\,J1627}
\newcommand{\GPM}{GPM\,J\,1839\ensuremath{-}10}
\newcommand{\PSR}{PSR\,J\,0901\ensuremath{-}4046}
 \newcommand{\srcDM}{\ensuremath{36.541}}
\newcommand{\DMerror}{\ensuremath{0.005}}
\newcommand{\DMunit}{pc\,cm\ensuremath{^{-3}}}
\newcommand{\srcDist}{\ensuremath{1.5}}
\newcommand{\Disterror}{\ensuremath{0.5}}
\newcommand{\Distunit}{kpc}
\newcommand{\srcRM}{\ensuremath{-7}}
\newcommand{\RMerror}{\ensuremath{1}}
\newcommand{\RMunit}{rad\,m\ensuremath{^{-2}}}
\newcommand{\srcPlong}{\ensuremath{10496.5575}}
\newcommand{\srcPerr}{\ensuremath{0.0005}}
\newcommand{\srcP}{\ensuremath{2.9}}
\newcommand{\srcFlong}{\ensuremath{0.00095269329(9)}}
\newcommand{\srcFerr}{\ensuremath{X\times10^{Y}}}
\newcommand{\srcFdotlong}{\ensuremath{X\times10^{Y}}}

\newcommand{\srcalpha}{\ensuremath{-6.2}}
\newcommand{\srcalphaerr}{\ensuremath{0.6}}
\newcommand{\srcq}{\ensuremath{-1.9}}
\newcommand{\srcqerr}{\ensuremath{1.4}}

\newcommand{\nmwadetections}{33}
\newcommand{\nmktdetections}{33}
\newcommand{\ndetections}{66}
\newcommand{\srcPdot}{\ensuremath{1.3\times10^{-11}}}
\newcommand{\srcPdotts}{\ensuremath{3.9\times10^{-11}}} 
\newcommand{\srcPdoterr}{\ensuremath{1.3\times10^{-11}}}
\newcommand{\Pdotunit}{\,s\,s\ensuremath{^{-1}}}

\newcommand{\srcPorb}{\ensuremath{2,283}}
\newcommand{\srcPorberr}{\ensuremath{84}}

\newcommand{\srcBp}
{\ensuremath{1.2\times10^{15}}}
\newcommand{\srcBpts}
{\ensuremath{2.4\times10^{15}}}

\newcommand{\srcBpwd}{\ensuremath{4\times10^{9}}}
\newcommand{\srcBptswd}{\ensuremath{8.1\times10^{9}}}

\newcommand{\srcSLum}{\ensuremath{1.3\times10^{25}}}
\newcommand{\srcSLumts}{\ensuremath{6\times10^{25}}}

\newcommand{\srcSLumwd}{\ensuremath{2.4\times10^{30}}}
\newcommand{\srcSLumtswd}{\ensuremath{1\times10^{31}}}

\newcommand{\srctau}{\ensuremath{58}}
\newcommand{\srctauts}{\ensuremath{22}}

\newcommand{\srcRLum}{\ensuremath{10^{28}}}

\newcommand{\ergpers}{erg\,s\ensuremath{^{-1}}}
\newcommand{\flux}{erg\,cm\ensuremath{^{-2}}\,s\ensuremath{^{-1}}}

\def\arc{\mbox{$^{\prime\prime}$}}
\def\nh{\hbox{$N_{\rm H}$}}

\title{A 2.9-hour periodic radio transient with an optical counterpart}

\author[0000-0002-5119-4808]{N. Hurley-Walker}
\affiliation{International Centre for Radio Astronomy Research, Curtin University, Kent St, Bentley WA 6102, Australia}
\author[0000-0001-6114-7469]{S.~J.~McSweeney}\affiliation{International Centre for Radio Astronomy Research, Curtin University, Kent St, Bentley WA 6102, Australia}
\author[0000-0003-2506-6041]{A.~Bahramian}\affiliation{International Centre for Radio Astronomy Research, Curtin University, Kent St, Bentley WA 6102, Australia}
\author[0000-0003-2177-6388]{N.~Rea}\affiliation{Institute of Space Sciences (ICE), CSIC, Campus UAB, Carrer de Can Magrans s/n, E-08193, Barcelona, Spain}\affiliation{Institut d'Estudis Espacials de Catalunya (IEEC), Esteve Terradas 1, Edifici RDIT, Parc Mediterrani de la Tecnologia (PMT) Campus UPC, 08860 Castelldefels, Spain}
\author[0009-0003-0996-9176]{C.~Horv\'{a}th}\affiliation{International Centre for Radio Astronomy Research, Curtin University, Kent St, Bentley WA 6102, Australia} 
\author[0000-0002-1691-0215]{S.~Buchner}\affiliation{South African Radio Astronomy Observatory, 2 Fir Street, Black River Park, Observatory 7925, South Africa}
\author[0000-0001-9080-0105]{A.~Williams}\affiliation{International Centre for Radio Astronomy Research, Curtin University, Kent St, Bentley WA 6102, Australia}
\author[0000-0001-8845-1225]{B.~W.~Meyers}\affiliation{International Centre for Radio Astronomy Research, Curtin University, Kent St, Bentley WA 6102, Australia}
\author[0000-0002-1468-9668]{Jay Strader}\affiliation{Center for Data Intensive and Time Domain Astronomy, Department of Physics and Astronomy, Michigan State University, East Lansing, MI 48824, USA}
\author[0000-0001-8525-3442]{Elias Aydi}\affiliation{Center for Data Intensive and Time Domain Astronomy, Department of Physics and Astronomy, Michigan State University, East Lansing, MI 48824, USA}
\author[0000-0003-1814-8620]{Ryan Urquhart}\affiliation{Center for Data Intensive and Time Domain Astronomy, Department of Physics and Astronomy, Michigan State University, East Lansing, MI 48824, USA}
\author[0000-0002-8400-3705]{Laura Chomiuk}\affiliation{Center for Data Intensive and Time Domain Astronomy, Department of Physics and Astronomy, Michigan State University, East Lansing, MI 48824, USA}
\author[0000-0002-2801-766X]{T.~J.~Galvin}\affiliation{International Centre for Radio Astronomy Research, Curtin University, Kent St, Bentley WA 6102, Australia}\affiliation{CSIRO Space \& Astronomy PO Box 1130, Bentley, WA 6102, Australia}
\author[0000-0001-7611-1581]{F.~Coti Zelati}\affiliation{Institute of Space Sciences (ICE), CSIC, Campus UAB, Carrer de Can Magrans s/n, E-08193, Barcelona, Spain}\affiliation{Institut d'Estudis Espacials de Catalunya (IEEC), Carrer Gran Capit\`a 2--4, E-08034 Barcelona, Spain}
\author[0000-0003-3294-3081]{Matthew Bailes}\affiliation{Centre for Astrophysics and Supercomputing, Swinburne University of Technology, Mail H30, PO Box 218, VIC 3122, Australia}


\begin{abstract}
We present a long-period radio transient (\src{}) discovered to have an optical counterpart, consistent with a cool main sequence star of spectral type M3. The radio periodicity occurs at the longest period yet found, 2.9\,hours, and was discovered in archival low-frequency data from the Murchison Widefield Array (MWA). High time resolution observations from MeerKAT show that pulsations from the source display complex microstructure and high linear polarisation, suggesting a pulsar-like emission mechanism occurring due to strong, ordered magnetic fields. The timing residuals, measured over more than a decade, show tentative evidence of a $\sim$6\,yr modulation. The high Galactic latitude of the system and the M-dwarf star excludes a magnetar interpretation, suggesting a more likely M-dwarf / white dwarf binary scenario for this system. 
\end{abstract}


\section{Introduction}\label{sec:intro}

Transient phenomena are often associated with high-energy or extreme physics, and historically their serendipitous detection has led to new astrophysical advances. In the radio sky, coherent non-thermal emission processes in extreme environments produce transient sources across a range of timescales; the shortest have been mined for pulsars, resulting in the discovery of the millisecond-timescale Fast Radio Bursts \citep[FRBs;][]{2007Sci...318..777L, 2022A&ARv..30....2P}.
Slower-timescale (hours to months) coherent emission can be detected from explosive transients such as gamma-ray bursts \citep{2000ARA&A..38..379V} or beamed jets from blazars \citep{1995PASP..107..803U}.

Timescales of minutes to hours have historically been poorly probed.
Despite the discovery of GCRT\,J1745$-3009$ by \cite{2005Natur.434...50H}, a radio source that produced five 11-minute duration pulses repeating every 77\,minutes, no similar objects were found until the advent of the SKA precursors.
Now we are seeing an explosion of discovery, with the emerging class of ``long-period radio transients'' (LPTs) \footnote{Otherwise called ``ultra-long period transients'', ``ultra-long period objects'', etc.}.

By searching the GaLactic and Extragalactic All-sky Murchison Widefield Array \citep[MWA;][]{2013PASA...30....7T,2018PASA...35...33W} eXtended \citep[GLEAM-X;][]{2022PASA...39...35H} survey at 72--231\,MHz using a visibility differencing technique, \cite{2022Natur.601..526H} discovered the radio transient GLEAM-X\,J1627$-52$, which produced $\sim$45-Jy, 30--60\,s-wide, highly linearly-polarised pulses every 18\,minutes for three months in 2018, with a dispersion consistent with a Galactic origin. This was followed by the discovery of GPM\,J1839$-10$, which repeated every 22\,minutes, and unlike the previous two very transient sources, was found to be active for at least three decades
\citep{2023Natur.619..487H}. Using the Australian SKA Pathfinder (ASKAP) at 888\,MHz, \cite{2024NatAs.tmp..107C} have discovered a $P\sim$54-min transient ASKAP\,J1935$+2148$ which produces both broad, linearly-polarised pulses, and short, circularly-polarised bursts. \cite{2024arXiv240612352D} have also detected a single 2-minute pulse from ASKAP\,J1755$-25$, which phenomenologically resembles the other sources, but has not yet been observed to repeat. \cite{2024arXiv240707480D} have detected a radio source with $P\sim$421\,s, CHIME J0630$+25$; at a distance of 170\,pc, it is the closest LPT discovered to date, compared to (fairly uncertain) distances of 1--8\,kpc for the other sources. \cite{2024arxiv240811536d} have recently noted linearly-polarised periodic radio emission with a period of $\sim$2\,hr from a binary M-dwarf / WD system, ILT\,J1101$+$5521, interpreting it as the spin and orbital periodicity of a polar-like system. 

So far, optical association of the LPTs has been difficult, since they have been discovered at low Galactic latitudes in crowded fields. For a few LPTs deep optical and infrared observations were reported. GLEAM-X\,J1627$-52$ was observed deeply with large facilities, but its imprecise radio localisation and position in crowded Galactic fields meant that none of the coincident sources could be unambiguously associated, with resulting optical and infrared limits of $g>23.7$, $i>22.8$ and $K>18.2$ magnitude \citep{2022ApJ...940...72R}. On the other hand, for GPM\,J1839$-10$, the 10\,m GranTeCan telescope could pinpoint a possible counterpart at its refined radio position with $K_s=19.7\pm0.2$, identified as a main-sequence mid-K or mid-M star (see Methods section in \citealt{2023Natur.619..487H}). As for the 54-min source ASKAP\,J1935$+2148$, limits of $g>23.3$, $i>23.1$, and $y>21.3$ have been derived \citep{2024NatAs.tmp..107C}. Only for ILT\,J1101$+$5521 \citep{2024arxiv240811536d} an optical counterpart could be conclusively pinpointed thus far. In this paper, we present a new LPT, \src{}, which has a clear optical counterpart identified with a low-mass star, and might exhibit two long periodicities.

\section{Observations and Data Analysis}\label{sec:obs}

\subsection{Identification}

Using the MWA, the GLEAM-X survey has observed the sky south of declination $+30^\circ$ in a series of $\sim$100~drift scans across 72--231\,MHz. 
Taken at a native time resolution of 0.5\,s in two-minute duration snapshots, the data are averaged to 4-s resolution to reduce processing costs. Each snapshot is imaged \citep[with \textsc{WSClean;}][]{2014MNRAS.444..606O} to form a deep model of the sky visibilities. Continuum-subtracted images are formed for each 4-s timestep; the resulting cube is then searched with various filters sensitive to pulsed, bursty, or variable emission. We detected the new transient source \src{} as a single 30-s wide pulse in a two-minute observation on 2018-02-04 at 14:57:02 UT, performed at 103--134\,MHz. Searching the archival MWA data back to the commencement of observatory operations in August~2013, we identified 33~further pulses. We found that the shortest interval between consecutive pulses was approximately three hours. Several long-duration drift scans were sensitive to the source's location, and no pulses were detected at shorter intervals.

\subsection{Radio follow-up}\label{sec:radio}

We performed follow-up observations with the MeerKAT radio telescope 
at UHF (550--1050\,MHz), using both correlator-mode at 2-s dump times and the Pulsars and Transients User Supplied Equipment \citep[PTUSE][]{2020PASA...37...28B}. In our longest observation, we observed \src{} from 2023-10-04 01:52:32 to 05:57:58, finding two distinct pulses with centroid times of arrival (TOAs) at 02:33:02 and 05:29:59, and no pulses at a shorter interval. The calibrator scans were chosen to be taken at non-integer divisors of the initial period, so that we would not miss any pulsations. We thus confirmed the initial periodicity estimate of $\sim3$\,hours.

MWA (at 170--200\,MHz) and MeerKAT (UHF) conducted a simultaneous observation of a single pulse at 2024-06-05 09:06:48. This enabled a constraint of the radio spectrum between 170 and 1050\,MHz, found to be best modelled by a curved power-law of the form:

\begin{equation}
    S_\nu = S_\mathrm{1\,GHz}\left(\frac{\nu}{\mathrm{1\,GHz}}\right)^\alpha \exp{q \left(\log{\frac{\nu}{\mathrm{1\,GHz}}}\right)^2}
    \label{eq:snu}
\end{equation}

 with $\alpha=\srcalpha{}\pm\srcalphaerr{}$ and $q=\srcq{}\pm\srcqerr{}$ (at a reference frequency of 1\,GHz). This pulse has a total radio luminosity of $2\Omega_\mathrm{1GHz} \left(\frac{D}{1.5\mathrm{kpc}}\right)^2\times 10^{29}$\,\ergpers{}, where $\Omega_\mathrm{1GHz}$ is the beam solid angle at 1\,GHz, and $D$ is the source distance (see \Sect~\ref{sec:distance}).

\begin{figure*}
    \includegraphics[width=\textwidth]{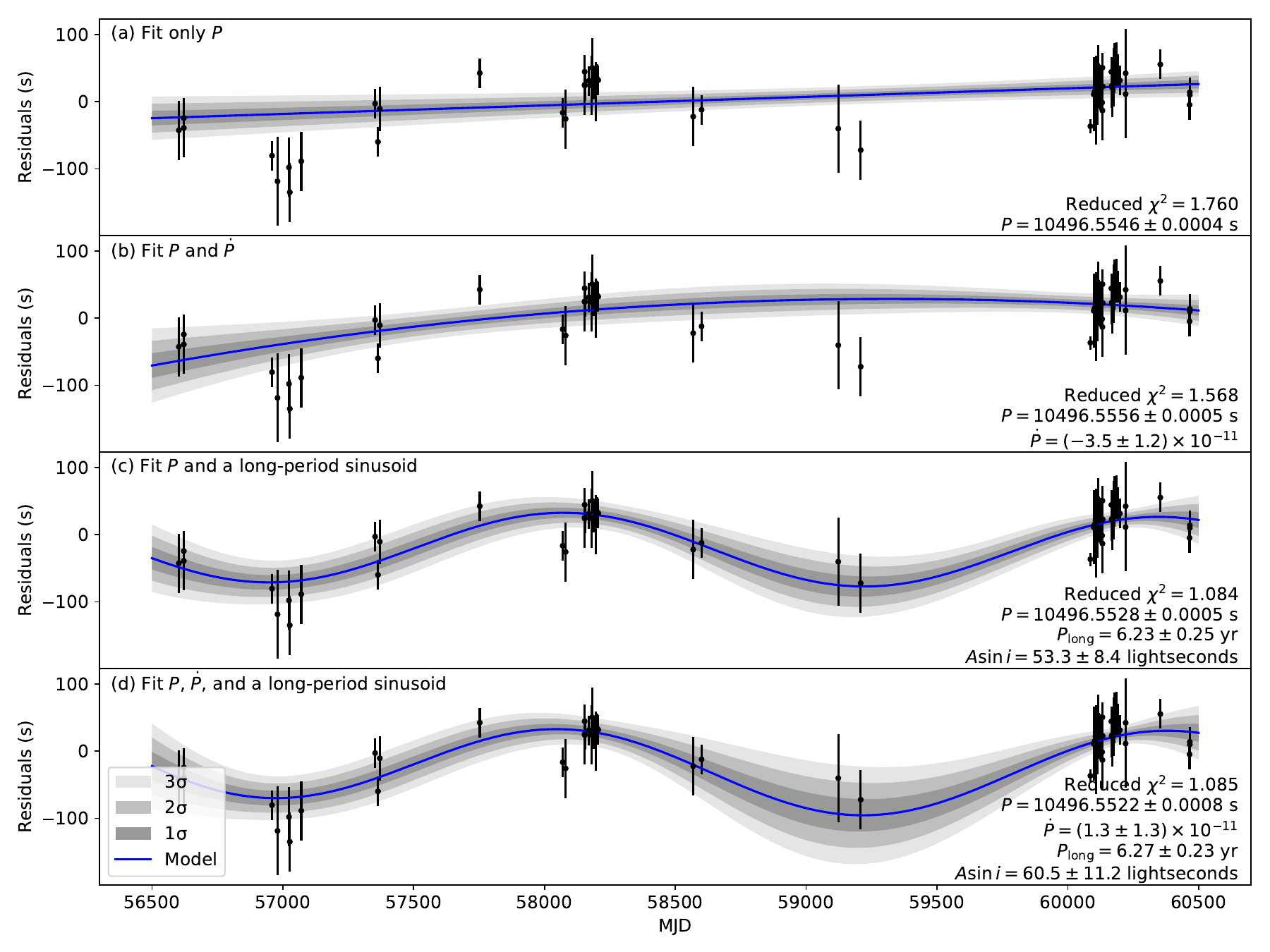}
    \caption{Various model fits to the timing residuals assuming an initial folding period of 10496.5531\,s. The fitted parameters are: $P$, the rotation period; $\dot{P}$, the spindown rate; $P_{\rm long}$, a sinusoidal period; and $A\sin i$, the sinusoidal amplitude projected onto the line of sight. Reported errors are 1-$\sigma$. The fit that includes only $P$ and $\dot{P}$ (b) produces a non-physical (negative) spin down. The fits with $P_{\rm long}$ (c \& d) are favoured over those without (a \& b).\label{fig:timing}}
\end{figure*}

The TOAs of individual pulses were determined using the same method described by \citet{2023Natur.619..487H}. The timing residuals (\Fig~\ref{fig:timing}) demonstrate an additional sinusoidal variation 
of the radio emitter. A timing analysis for 2013--2020 yields a period of $P=\srcPlong{}\pm\srcPerr{}$\,s, a period derivative of $<$\srcPdot{}\Pdotunit{}, and this additional modulation at $P_{\rm long}=$\srcPorb{}$\pm\srcPorberr{}$\,days. The spin-down rate $\dot{P}$ should be considered as an upper limit given that its error encapsulates zero spin-down (see bottom panel of \Fig~\ref{fig:timing}).

Following the method of \cite{2022MNRAS.514L..41E}, we choose a conservative $3\sigma$ upper limit of $\dot{P}<\srcPdotts{}$\,\Pdotunit{}, and thereby derive a beam solid angle of $\Omega_\mathrm{1GHz}<1.5\times10^{-5}$, consistent with the short pulse duty cycle ($\sim$30--60\,s in \srcP{}\,hr). This produces a radio luminosity limit of $\lesssim 3 \times10^{24}$\,\ergpers{}; pulses up to $20\times$ brighter are also observed. No secular change in brightness with respect to time has been observed, after correcting the MWA and MeerKAT data to the same frequency.

To measure the position, the MeerKAT correlator data at the time intervals covering the brightest seven pulses were imaged. The position measurements showed variance beyond that which would be expected from thermal noise, implying small ($\sim0\farcs5$) phase-referencing (astrometric) errors which vary between observations, consistent with expected variations from ionospheric refractive shifts. We thus derive the position from the average of the position measurements, and the error by the standard deviation: in J2000, RA$=$07$^\mathrm{h}$04$^\mathrm{m}$13.$^\mathrm{s}$19 and Dec$=-37^\circ$06$'$14\farcs3, with a (1-$\sigma$) uncertainty of 0\farcs3.

No persistent radio source was detected down to a 3$\sigma$ limit of 30\,$\mu$Jy\,beam$^{-1}$ using MeerKAT UHF data integrated outside of the pulse activity windows.

\begin{figure*}
    \centering
    \includegraphics[width=\textwidth]{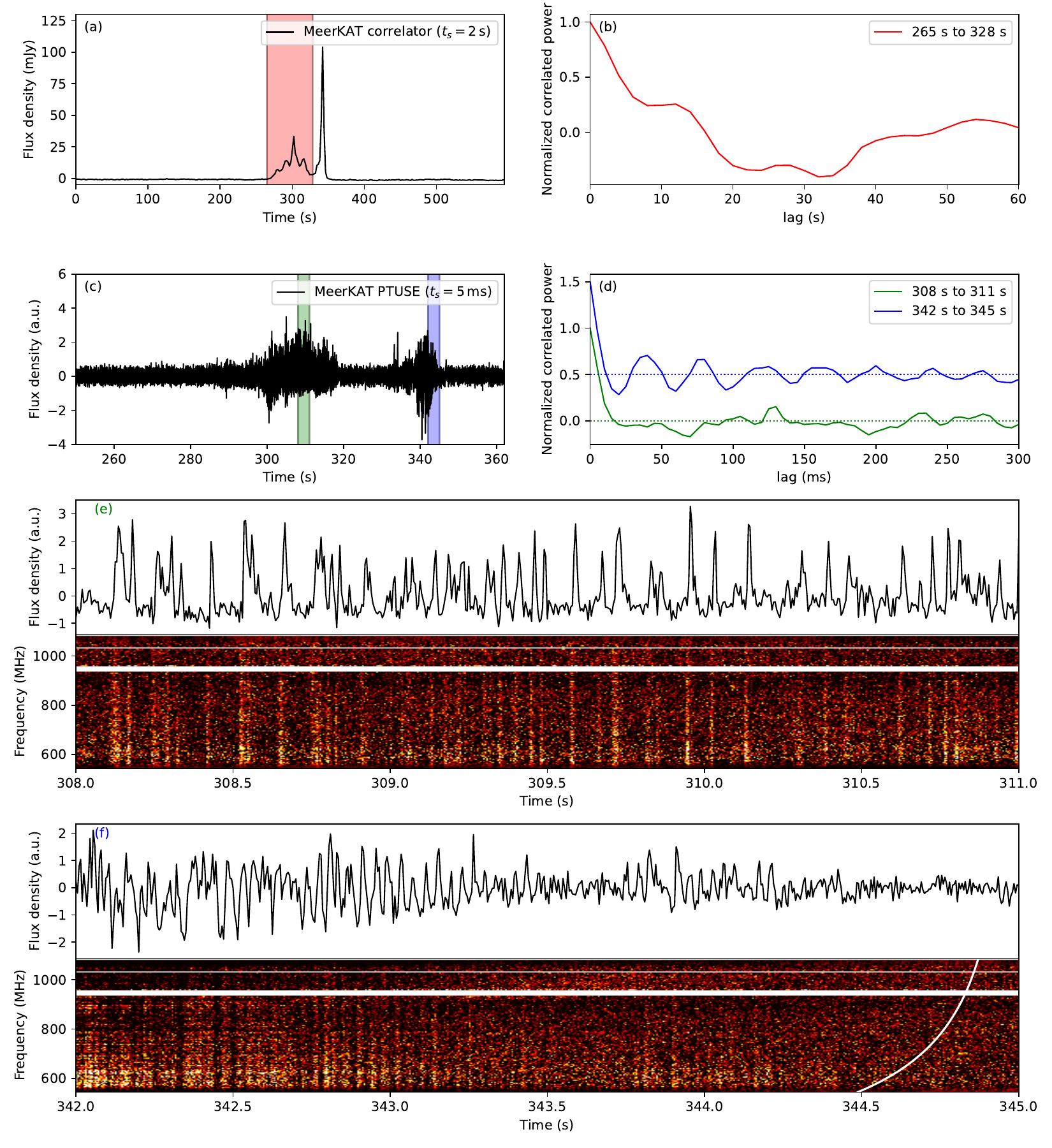}
    \caption{Autocorrelation analysis of both the correlator lightcurve sampled at 2\,s (a \& b) and the PTUSE lightcurve sampled at 5\,ms (c \& d), and subsets of the PTUSE data with significant microstructure (e \& f). The autocorrelations in (b) \& (d) include only the corresponding data highlighted in panels (a) \& (c); a vertical offset has been applied to the blue curve in panel (d) for visual clarity. The preprocessing of the PTUSE data (see main text) has effectively removed variations on timescales $\gtrsim 0.4\,$s, but retains finer time structures. In at least one section (colored blue, with detail shown in panel (f)), a quasi-periodicity of ${\sim}40\,$ms is clearly evident. The dynamic spectra have been dedispersed to \srcDM{}\,\DMunit{}. In panel (f), a white curve indicates how a zero-DM transient signal would appear.}
    \label{fig:microstructure}
\end{figure*}

Smaller timescales were probed with the PTUSE data, which has a native time resolution of ${\sim}10\,\mu$s. We discovered abrupt changes in the brightness in both the precursor emission (at ${\sim}300\,$s in panel (a) of Fig. \ref{fig:microstructure}) as well as in the bright peak (at ${\sim}340\,$s). Such changes were predominantly in the form of narrow peaks, $\lesssim 10\,$ms in width, but also occasionally appear as similarly sized \emph{reductions} in brightness (in particular panel (c) of \Fig~\ref{fig:microstructure}). Although the PTUSE data also showed some instrumental gain changes (whose origin remains unclear), these could be distinguished from the abrupt brightness changes of \src{} by virtue of the astrophysical dispersion of the intrinsic emission, corresponding to a dispersion delay of $\tau_{\rm DM} \approx 0.4\,$s across the observing band. This microstructure enabled a highly accurate calculation of the dispersion measure: \srcDM{}$\pm$\DMerror{}\,\DMunit{}. 

After correcting for dispersion and frequency-scrunching, the resulting light curve, shown in panel (c) of Fig. \ref{fig:microstructure}, has effectively had all instrumental gain changes (at all timescales) removed, and retains only brightness variations of \src{} smaller than $\tau_{\rm DM}$.
The ACF analysis of the cleaned PTUSE lightcurve (and a few representative subsections of it) is shown in panel (d). Only in one section, corresponding to the trailing side of the brightest peak, did we find clear evidence of a quasi-periodicity of ${\sim}40\,$ms; elsewhere, we did not detect any significant periodicity.

The radio polarisation behaviour of the source is extremely complex, with strong linear (20--50\,\%) and some circular polarisation (10--30\,\%) that changes on short (ms) timescales. This behaviour is in contrast with the high (95\,\%) linear polarisation and flat polarisation angles observed in \GLX{}, and bears some similarity to the complex behaviour recently uncovered in \GPM{} (Men et al. 2024, in press). 
A companion paper, McSweeney et al. in preparation, discusses the polarisation behaviour of all three sources in more detail. For those parts of the brightest pulses where the Faraday rotation is clear, we derive a rotation measure of $\srcRM{}\pm\RMerror{}$\,\RMunit{}.

\subsection{Optical \& Infrared counterpart}\label{sec:optical}

\begin{figure*}
\centering
\hspace{-0.5cm}
\includegraphics[width=7.5cm]{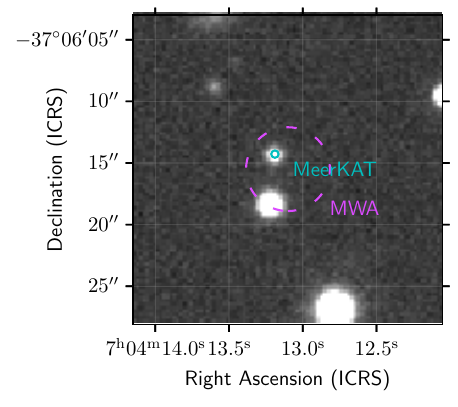}
\includegraphics[width=10.5cm]{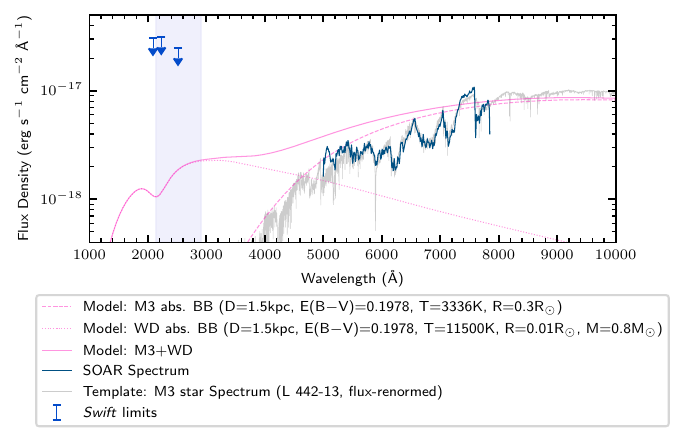}
\caption{Left: DECam r-band image of the vicinity of the radio transient, with its 1-$\sigma$ localization regions by MWA (dashed magenta circle) and MeerKAT (solid cyan circle) overplotted. Right: SOAR (4m-class) low-resolution optical spectrum of the optical counterpart (blue solid line) overplotted on an example M3V star (grey line, flux and extinction renormalized to match). The \textit{Swift}-UVOT limits are shown with downward-pointing arrows. An AR~Sco-like WD with $T=11,500$\,K, $M=0.8$\,M$_\odot$ is shown with the dotted pink line, and the summed WD+M3V spectrum is shown with a solid pink line. \label{fig:optical}}
\end{figure*}

We inspected and analyzed archival images and catalogs covering the region, including images by the Dark Energy Camera at the Blanco 4\,m telescope at the Cerro Tololo Inter-American Observatory in Chile \citep{2015AJ....150..150F}, and catalogs from Gaia Data Release 3 \citep[Gaia~DR3;][]{2023A&A...674A...1G} and the VISTA Hemisphere Survey \citep[VHS;][]{2013Msngr.154...35M}. We identified an optical counterpart 0\farcs3 away from the MeerKAT localization (\Fig~\ref{fig:optical}; left) with Gaia designation of Gaia~DR3~5566254014771398912 and coordinates of RA$=$07$^\mathrm{h}$04$^\mathrm{m}$13.$^\mathrm{s}$19, Dec$=-37^\circ$06$'$14\farcs58, with uncertainties of 0.8 and 1 milliarcseconds in RA and Dec, respectively. Considering the stellar density in this region in Gaia~DR3, we estimated the probability of a chance alignment with a separation of $\leq$0\farcs3 to be $0.05\%$. 

We performed spectroscopy of this optical counterpart with the Goodman Spectrograph \citep{2004SPIE.5492..331C} on the SOAR telescope on 2023 October 25 with a 400\,line\,mm$^{-1}$ grating and a 1.2\arcsec-longslit, yielding a FWHM resolution of 6.7\,\AA\ over a wavelength range of 4000--7840\,\AA. Two 30-min exposures were taken, for a total exposure time of 1\,hr. The data were reduced, optimally extracted, and combined in the normal manner in \textsc{IRAF} \citep{1986SPIE..627..733T} and the spectra were flux-calibrated using a spectrum of the spectrophotometric standard EG 274 via the IRAF tasks \texttt{standard}, \texttt{sensfunc}, and \texttt{calibrate}.

The SOAR spectrum shows significant molecular lines, such as TiO absorption features (\Fig~\ref{fig:optical}; right), consistent with M stars. Comparing with PHOENIX spectral models \citep{2013A&A...553A...6H} and X-shooter spectral library DR3 \citep{2022A&A...660A..34V}, we inferred a spectral type of M3 with a temperature of $\sim3400$~K. Combined with the constraints on the distance (see \S\ref{sec:distance}) and apparent brightness of the source, which rules out a red giant, we conclude that the optical source is an M3V star. In \Fig~\ref{fig:optical} (right) we also show the comparison with a WD+M3V system with a WD with $T=11,500$\,K and $M=0.8$\,M$_\odot$ (like AR~Sco), showing that such a WD star would not be detectable (see \Sect~\ref{sec:wd} for a more thorough treatment of temperature upper limits).

%

To search for UV emission from a potential WD, we observed \src{} with Swift/UVOT \citep{2004ApJ...611.1005G, 2005SSRv..120...95R} for 4ks on 2023-06-15 (Obs. ID. 00016076001) with UVOT in 0x308f mode (equal weighting of UV filters). We reduced and analyzed the data following standard procedures.\footnote{\url{https://www.swift.ac.uk/analysis/uvot/index.php}} We found no evidence for UV emission at the location of the source, with 3-$\sigma$ Vega magnitude upper limits of UM2 = 20.4 , UW1 = 20.5 , UW2 = 20.6. These upper limits are plotted in Figure~\ref{fig:optical}.

\subsection{X-ray observations}\label{sec:xray}

We observed \src{} using XMM-Newton beginning on November 16, 2023, at 23:07 UT and ending on November 17, 2023, at 09:02 UT.  This time span overlapped with a 7-hour dwell starting on November 17th at 01:20 UT on the MeerKAT telescope (in UHF). All European Photon Imaging Camera (EPIC) cameras were set to Full-Frame mode.

The data were processed using the Science Analysis System (\textsc{sas} version 21.0; \citealt{gabriel04}). The 
data were significantly affected by strong background flares. After removing these flares, the usable data was reduced to $\simeq$15.2\,ks for EPIC-pn and $\simeq$16.2\,ks for the EPIC-MOSs. No X-ray source was detected at the position of \src{}. We estimated the 3-$\sigma$ upper limit on the EPIC-pn net count rate at the source position to be 2$\times$10$^{-3}$\,counts\,s$^{-1}$ over the 0.3–10\,keV energy range. This results in a 3-$\sigma$ upper limit on the unabsorbed flux of $\sim9\times10^{-15}$\,erg\,s$^{-1}$\,cm$^{-2}$ assuming a power law spectrum with a photon index of $\Gamma=2$, or $\sim5\times10^{-15}$\,erg\,s$^{-1}$\,cm$^{-2}$ assuming a black body spectrum with a temperature of kT=0.3\,keV. 
In both cases, we accounted for the effects of absorption by the interstellar medium by incorporating the \textsc{tbabs} absorption component \citep{2000ApJ...542..914W} into the model, adopting a hydrogen column density of $N_H = 2.7\times10^{21}$\,cm$^{-2}$ (this is the expected value along the line of sight to the source within our Galaxy; \citealt{willingale13}). Combined with the constraints on the distance (see \S\ref{sec:distance}) these limits convert to a luminosity of L$_{\rm X} < 2.2\times10^{30}$\,erg~s$^{-1}$ for the assumed power-law spectrum and L$_{\rm X} < 1.1\times10^{30}$\,erg~s$^{-1}$ for the black body case.

During the time XMM-Newton/EPIC was on source, two radio bursts were detected with MeerKAT, at 2023-11-17T02:01:25 and 04:56:29 (UTC). We shifted the X-ray and radio time stamps to the Solar System's barycenter, using the radio position and the DE-430 JPL ephemeris for reference, and extracted images from various time intervals around these bursts. No excess emission was detected above the background level. 
We measured upper limits on the EPIC-pn net count rate within a 100-s time window centered on the times of the two bursts, yielding values of 0.09\,counts\,s$^{-1}$ and 0.06\,counts\,s$^{-1}$, respectively (3-$\sigma$; 0.3--10\,keV). Assuming the same spectral models as those used for the average emission, the corresponding luminosity limits are L$_{\rm X, PL} < 1.1 \times10^{32}$\,erg~s$^{-1}$ and L$_{\rm X, BB} < 6.6\times10^{31}$\,erg~s$^{-1}$ for the first burst and L$_{\rm X, PL} < 7.3\times10^{31}$\,erg~s$^{-1}$ and L$_{\rm X, BB} < 4.4\times10^{31}$\,erg~s$^{-1}$ for the second burst. Note that all luminosities are dependent on the distance $D$ via $\left(\frac{D}{1.5\mathrm{kpc}}\right)^2$, which currently has significant uncertainty (see below).


\subsection{Distance}\label{sec:distance}

We can convert our accurate estimate of the dispersion measure (\Sect~\ref{sec:radio}) into a distance using electron density models of the Milky Way. Depending on the model used, the resulting estimated distance is $0.4(\pm0.1)$ \citep[YMW16;][]{2017ApJ...835...29Y} or $1.8(\pm0.5)$\,kpc \citep[NE2001;][]{2004ASPC..317..211C}. 

We may also estimate the distance using the parallax measured by the \textit{Gaia} mission \citep{2016A&A...595A...1G}. Gaia~DR3 reports a parallax of $-0.22\pm1.0$ mas, indicating that any inference of distance would be heavily reliant on prior models \citep[e.g.,][]{2018arXiv180407766H}. We estimated the distance using the \cite{2018AJ....156...58B} geometric model based on stellar distributions and found a value of 2.5$_{-0.9}^{+1.5}$~kpc. Considering the optical spectrum, and comparing with dereddened spectra of nearby red dwarfs, we estimated $E(B-V)=0.2$, which in the context of Galactic extinction maps \citep{2018JOSS....3..695M,2024A&A...685A..82E} suggests a distance of $\gtrsim1$~kpc. Combining these constraints, we assume a distance of $1.5\pm0.5$\,kpc throughout the remainder of the paper, but we note that this is significantly uncertain.

\section{Discussion}\label{sec:discussion}

The discovery of linearly-polarised, $P = 2.9$-hour pulses from \src{} places it in the burgeoning class of ``long-period radio transients''. We have identified an optical counterpart for an LPT, with a clear M3-dwarf spectrum. The radio luminosity of the periodic bursts are as high as $\sim10^{26}$\,erg\,s$^{-1}$,
while at X-ray energies the bursts are found to be fainter than L$_{\rm X}\sim10^{32}$\,erg\,s$^{-1}$. 

The radio emission shows microstructures and a peculiar long-period timing modulation of $\sim$6\,yr. It is tempting to interpret $P_{\rm long}$ as an orbital period of a system composed of an M-dwarf and a 2.9\,hr spinning NS or a WD (see also below). However it is possible that the long-term variation in timing residuals is due to a stochastic, red process analogous to timing noise in canonical pulsars, whose physical origins are still unknown. Assuming a simple period-only model, the amplitude of our residuals is $\pm50$\,s, i.e. an amplitude of 1\,\% of the spin period, and only a factor of a few larger than the pulse width. If we assume most of the power in the residuals is near the best-fitting frequency of $1/P_{\rm long} \approx 0.16\,{\rm yr}^{-1}$, then we can compare, to first order, the implied power spectral density of the TOA residuals to the pulsars studied by \citet{2019MNRAS.489.3810P}, who characterise timing noise as a power-law power spectrum. We find that the power spectral density of \src{} significantly exceeds that of the pulsars in their sample (extrapolated to the same frequency). It remains unclear how (or even if) \src{}'s vastly larger period might affect this kind of timing noise.

It is also worth noting that ascribing the absolute peak-to-peak amplitude of $\sim$100\,s to a change in the compact object's angular momentum implies a phenomenal transfer of energy, regardless of whether the compact object is a NS or a WD. Such a significant energy transfer seems improbable, suggesting that the observed variations more likely arise from changes in the emission region rather than rotational glitches in the compact object itself.

Below we discuss several possibilities starting from the firm detection of periodic radio pulses and the M-dwarf optical spectrum of the counterpart.

\subsection{Stellar radio flares or exo-planet interaction?}

The presence of an M-dwarf coincident with the radio emission is unseen in other long-period radio transients.
Radio emission from stars is typically circularly polarized, and appears in flares or bursts due to sunspot activity \citep{1985ARA&A..23..169D, 2017ApJ...836L..30L} or, potentially detectable at low radio frequencies ($<$200\,MHz), star--exoplanet interactions \citep{2020NatAs...4..577V}. 
Such flares would be wholly undetectable at $10^2$--$10^3$\,pc, and the strong periodicity is also difficult to explain with a purely stellar origin. The closest analogue is CU~Virginus, a hot, magnetic chemically-peculiar, Ap-type star 72\,pc away, which has been observed to produce pulses of 100\,\%-circularly-polarised emission at 1--3\,GHz, each lasting an hour, and repeating every 12\,hours, the rotation rate of the star \citep{2012MNRAS.421.3316L}. These are thought to be produced from electron cyclotron maser emission in the magnetosphere of the star. The high linear polarisation fraction (up to $50$\,\%) of \src{} would be difficult to explain with this emission mechanism. \cite{2021ApJ...921....9D} observed pulsations from CU~Virginus of brightnesses up to 100\,mJy at 1--2\,GHz; even these are $\sim$400 times less luminous than the brightest pulses from \src{}.

We therefore exclude the possibility that the M-dwarf alone produces the radio emission, and conclude that it is in a binary orbit with an as-yet optically undetected radio emitter.

\subsection{A binary system with a neutron star}

Since the optical spectrum shows no evidence of a further main-sequence companion (\Fig~\ref{fig:optical}), we suggest that the radio emitter is a compact stellar remnant. The high linear polarisation suggests ordered magnetic fields. Therefore, the companion is likely to be an NS or WD. In both cases, periodic radio emission might be generated, although not all the observational properties of \src{} are easily interpreted.

In the NS case, the high Galactic latitude of the system ($|b|=13^\circ$), the M-dwarf companion and the faint X-ray emission argue against a magnetar companion. These young NSs are generally isolated, often lie within massive star clusters or supernova remnants, and field decay exponentially dissipates their surface magnetic energy on Myr timescales \citep{2013MNRAS.434..123V}. This time is too short to allow them to leave the plane of the Galaxy even for extreme proper motions. If this source is a NS / M-dwarf binary, then the NS is necessarily an old pulsar with a relatively low $B$ field. Many such systems are known in different evolutionary stages, from radio-emitting millisecond pulsars \citep{2008LRR....11....8L} to accreting low-mass x-ray binaries \citep{2023hxga.book..120B} or transitional systems \citep{2022ASSL..465..157P}. However in all such cases the pulsar has undergone a recycling period due to accretion from the companion star spinning very fast, with usually hours-timescale orbits. For an NS pulsar binary to have a 2.9\,hr spin period while still maintaining enough magnetic field strength or rotational energy to produce radio pulsations, the NS must have been accreting in the propeller regime for an extended period. This process would gradually slow its rotation while preserving its magnetic field, but would require very fine-tuned conditions. Even then, the magnetic field necessary to power the current radio emission would exceed 10$^{17}$\,G, which is unrealistically large for an aged NS required for this scenario. The spin-down luminosity $L_\mathrm{spin}\lesssim4.4\times10^{23}$\,\ergpers{} would still be at least an order of magnitude too low to explain the radio luminosity.  Therefore, if this system is indeed a NS binary with a 2.9-hour spin period, it would challenge our current understanding of NS accretion and magnetic field evolution.
Alternatively, interpreting the 2.9-hour periodicity as an orbital period introduces more challenges, since the radio transient nature of the system, micro-structures and small duty cycle are difficult to explain through geometric considerations.

\cite{2024NatAs...8..230K} analysed the quasi-periodicity microstructure timescale $P_\mu$ of NS pulsations across six orders of magnitude of rotational period, finding a universal relation of $P_\mu = A_\mathrm{P} P^\gamma$\,ms, with  $A_\mathrm{P}=1.12\pm0.14$ and $\gamma=1.03\pm0.04$. This would predict $P_\mu$ = 15.5\,s for \src{}. Most pulses detected with MeerKAT are a maximum of 80\,s wide without much distinct substructure visible above the noise. In \Fig~\ref{fig:microstructure} we present the brightest detected pulse, and an auto-correlation analysis on data from both the correlator (timescales 2\,s--120\,s) and PTUSE (1\,ms--400\,ms). Intermediate timescales were contaminated with instrumental artefacts, likely from the PTUSE gain calibration. On the longer timescales, we see a peak in the ACF at 12\,s, broadly consistent with the \citeauthor{2024NatAs...8..230K} prediction.

On shorter timescales, the observed peak at $40\,{\rm ms} \approx 4 \times 10^{-6}\,$periods is much shorter than the \cite{2024NatAs...8..230K} relation, but can be compared to the spacing of nanoshots observed in the Crab pulsar, $\lesssim 0.1\,\mu{\rm s} \approx 3 \times 10^{-6}\,$periods \citep[cf.][Fig. 3]{2007ApJ...670..693H}. However, examples of known nanostructure are rare, and there is no universal relation for nanostructure analogous to the microstructure relation discussed above, so it would be premature to conjecture that the physical cause of the fine structure of \src{} is in any way related to that of the Crab nanoshots.
The behavior of the fine structure is also comparable to that of PSR\,J0901$-$4046 \citep[][esp. Extended Data Fig. 2]{2022NatAs...6..828C}, whose ACFs indicate quasi-periodicities on similar timescales (i.e. tens of milliseconds), even though the ratio of the quasi-periodic timescale to the rotation period is two orders of magnitude larger than for \src{}.

\subsection{A binary system with a white dwarf}\label{sec:wd}

Another possibility is that the compact object in the binary system is a WD, presumably with a strong magnetic field. The spin-down luminosity in this scenario would be $\lesssim4.4\times10^{29}$\,\ergpers{}, potentially more easily driving the observed radio emission. In polars, a type of magnetic cataclysmic variable \citep[MCV; see ][for a review]{1995cvs..book.....W}, the magnetic field of the white dwarf is 10--200\,MG, and the two stars are locked in synchronous rotation. Intermediate polars (IPs) have slightly weaker magnetic fields (1--10\,MG); the WD rotates at a faster rate than the orbit. With an even faster WD rotation rate, AR~Sco-type systems consist of an M-dwarf/WD pair in a tight ($\sim4$-hr) orbit, such that the M-dwarf is nearly Roche-lobe filling  \citep{2016Natur.537..374M}. Interactions between the two stars cause pulsar-like radio emission from the magnetosphere of the WD, which has been directly observed in J1912-44, as the beam of the WD crosses our line-of-sight \citep{2023NatAs...7..931P}. The two known ``WD pulsar'' systems were selected by looking for high optical variability, which preferentially selects for nearby systems with quickly-spinning WDs ($P_\mathrm{spin}\sim$minutes). It is therefore possible that slowly-rotating AR~Sco-type systems could have been missed by optical searches. The observed periodicity of $\srcP{}$\,hr {might be consistent with both the rotation rate and/or the orbit of the WD in the MCV scenario} \citep{2004ApJ...614..349N}.

\begin{figure}
    \includegraphics[width=\linewidth]{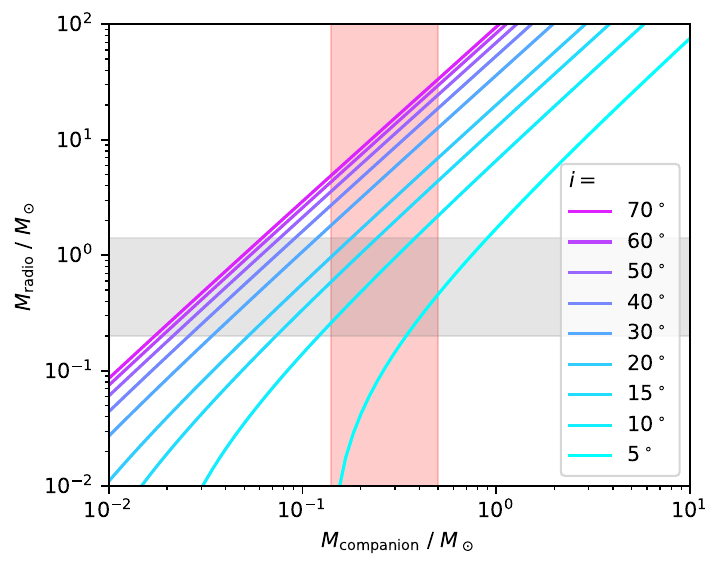}
    \caption{Allowable inclination angles $i$ for a circular orbit with period $P_{\rm long}=\srcPorb{}$\,day of an unseen radio emitter and the visible M3 dwarf. The grey shaded area shows the mass range of magnetic white dwarfs \citep{2023ApJ...944...56A} and the red shaded area shows the mass range of M3V dwarfs \citep{2020A&A...642A.115C}. For an MCV interpretation, a low inclination angle is preferred.\label{fig:massratio}}
\end{figure}

In the hypothesis of the long $\sim6$\,yr periodicity being the orbit of the system, the small amplitude of the timing residuals ($\pm$50s) implies a very low inclination angle (\Fig~\ref{fig:massratio}). The beam opening angle of this system would be extremely small, just $\sim$30\,s$/10,000\,\mathrm{s} \sim 0.6$\%. No MCVs have been ever observed with such long orbital period, although long period binary WD systems exist \citep{Ciardullo1999}.

The recently discovered ILT\,J1101$+$5521 \citep{2024arxiv240811536d}, a $\sim2$\,hr synchronized polar system comprising a WD and an M-dwarf showing periodic and coherent radio emission, might be a twin system to \src{}. In this scenario, the 2.9\,hr periodicity might be interpreted as both the spin and the orbital period, and the $\sim6$\,yr modulation either as timing noise or as a possible super-orbital modulation. The polarisation properties of this system are similar to \src{}, with a mix of linear and circular polarisation, which could be explained by a relativistic version of the electron cyclotron maser mechanism \citep{2024arXiv240905978Q}.\footnote{The polarisation properties of AR~Sco published by \cite{2017NatAs...1E..29B} demonstrate the high-frequency radio properties of the interaction between the WD and the M-dwarf, rather than the WD pulsar beam, and are therefore not directly comparable. \cite{2023NatAs...7..931P} do not describe the radio polarisation properties of the WD pulsar beam of J\,1912$-44$.}
We note that an orbital period of 2.9\,hr lies in the orbital period gap for MCVs \citep{Knigge2011}, something that have been observed for some polars, not for IPs. On the other hand, if interpreted as the spin period of an IP with a longer orbital period, this would represent the slowest IP ever discovered. We therefore tentatively suggest that \src{} might be a polar system, similar to ILT\,J1101$+$5521.
Assuming an M3V stellar mass of 0.32$M_\odot$ and a WD mass of 0.8$M_\odot$, the Roche lobe radius of the M3V star is 0.2\,A.U. For a 2.9\,hr circular orbit with identical masses, the orbital radius would be 0.005\,AU, placing the WD well within the Roche lobe.

 The UV limits derived by a 4\,ks of \textit{Swift} observations yielded
upper limits on the UV/blue emission, shown in \Fig~\ref{fig:optical}. This rules out the presence of a hot WDs with $T\gtrsim$20,000\,K, i.e. about 20\% of the magnetic WD population. Furthermore, no strong emission lines are seen in the infrared spectrum, which could be expected for a hot WD in a close orbit. However, the WDs in J1912-44, AR~Sco and ILT\,J1101$+$5521 are all much cooler, with the latter having a temperature of $T_{\rm eff}\sim5156$\,K \citep{2024arxiv240811536d}. The upper limit we derive from the infrared spectrum and the UV observations on the presence of a WD in the system are  not strongly constraining. Sensitive observations below a wavelength of 400\,nm are required to determine whether a WD is present.

\section{Conclusion}

We have discovered a new long-period radio transient with the longest period yet detected (2.9\,hr), and an optical counterpart with an M-dwarf spectrum. In this discovery paper we argue that:
\begin{enumerate}
    \item The radio emission does not arise purely from the M-dwarf, and the system is most likely a binary with an unseen radio emitter, probably a neutron star or relatively cool magnetic white dwarf;
    \item A binary system with an M-dwarf/neutron star is disfavoured by several arguments. If the radio emission is produced by a strong magnetic field, the hosted NS should be young, which is disfavoured by the high Galactic latitude. On the other hand, a scenario with a low-field NS in a low-mass binary (like the transitional pulsars or recycled pulsars), with the radio emission being dipolar spin-down powered, is disfavoured since the radio luminosity is several orders of magnitude larger than the limits on the spin-down power. 
    \item A binary system with an M-dwarf/white dwarf is a viable scenario. The radio emission in this case should be generated by a stellar wind flowing from the M-dwarf companion onto the magnetosphere of the white dwarf, where it is accelerated, producing radio emission. If this can be confirmed, these systems can be evolutionarily connected to AR~Sco-type systems, intermediate polars, and polar WDs. The 2.9\,hr is most likely the spin and the orbit of a polar system, although other configurations cannot yet be completely excluded.  A similar system might be the radio emitting polar candidate ILT\,J1101$+$5521 \citep{2024arxiv240811536d}.
    \item The microstructure in the radio pulses is consistent with the relationship seen in neutron stars \citep{2024NatAs...8..230K}, but finer-timescale ($\sim$40-ms) nanostructure is observed, whose origin is unknown; it is still unclear how these microstructures behave in radio-emitting WD binaries. 
    \item There is tentative evidence for a $\sim6$-year periodicity possibly due to timing noise. If interpreted as an orbital period, it would imply that the system components are separated by a distance greater by an order of magnitude than that which is seen in AR~Sco-type or ILT\,J1101$+$5521 systems, making similar interactions unlikely. More data are needed to disentangle the nature of this long-timescale periodicity.
\end{enumerate}

To conclusively determine the nature of this system, we suggest further radio monitoring to constrain the timing, sensitive blue/UV observations to search for a WD, and a measurement of the radial velocity of the M-dwarf by studying Doppler shifts of its spectral lines. Determining the nature of this system would be a further step in assessing the true nature of the enigmatic long-period radio transients.

\begin{acknowledgements}
\small
 We thank the anonymous referees for their comments, which improved the quality of this paper.
 This scientific work uses data obtained from Inyarrimanha Ilgari Bundara / the Murchison Radio-astronomy Observatory. We acknowledge the Wajarri Yamaji People as the Traditional Owners and native title holders of the Observatory site.
 This work was supported by resources provided by the Pawsey Supercomputing Research Centre’s Setonix Supercomputer (https://doi.org/10.48569/18sb-8s43).
 The Murchison Radio-astronomy Observatory and the Pawsey Supercomputing Research Centre are initiatives of the Australian Government, with support from the Government of Western Australia and the Science and Industry Endowment Fund.
Support for the operation of the Murchison Widefield Array is provided by the Australian Government (NCRIS), under a contract to Curtin University administered by Astronomy Australia Limited.
The authors would like to thank SARAO for the approval of the MeerKAT DDT request DDT-20230529-NH-01. The MeerKAT telescope is operated by the South African Radio Astronomy Observatory, which is a facility of the National Research Foundation, an agency of the Department of Science and Innovation (DSI).
 This research is based on observations obtained with \emph{XMM-Newton}, an ESA science mission with instruments and contributions directly funded by ESA Member States and NASA. We thank N.~Schartel for approving our DDT request and the \emph{XMM-Newton} Science Operations Centre for carrying out the observation. 
Based in part on observations obtained at the Southern Astrophysical Research (SOAR) telescope, which is a joint project of the Minist\'{e}rio da Ci\^{e}ncia, Tecnologia e Inova\c{c}\~{o}es (MCTI/LNA) do Brasil, the US National Science Foundation’s NOIRLab, the University of North Carolina at Chapel Hill (UNC), and Michigan State University (MSU).
 N.H.-W. is the recipient of an Australian Research Council Future Fellowship (project number FT190100231). N.~R. is supported by the European Research Council (ERC) via the Consolidator Grant ``MAGNESIA'' under grant agreement No. 817661 and from an ESA Science Faculty Visitor program to ESTEC (funding reference ESA-SCI-E-LE-054). F.~C.~Z. is supported by a Ram\'on y Cajal fellowship (grant agreement RYC2021-030888-I). N.~R. and F.~C.~Z acknowledge support from the Catalan grant SGR2021-01269 and by the program Unidad de Excelencia Mar\'ia de Maeztu CEX2020-001058-M. C.~H. acknowledges a Doctoral Scholarship and an Australian Government Research Training Programme scholarship administered through Curtin University of Western Australia. J.S. acknowledges support from the Packard Foundation and National Science Foundation (NSF) grant AST-2205550. L.C. is grateful for support from NSF grants AST-2107070 and AST-2205628 and NASA grant 80NSSC23K0497.
E.A. acknowledges support by NASA through the NASA Hubble Fellowship grant HST-HF2-51501.001-A awarded by the Space Telescope Science Institute, which is operated by the Association of Universities for Research in Astronomy, Inc., for NASA, under contract NAS5-26555.

\end{acknowledgements}

\bibliography{refs}{}
\bibliographystyle{aasjournal}




\end{document}